\documentclass[conference]{IEEEtran}
\usepackage{amsmath,amssymb,amsfonts}
\usepackage{hyperref}
\hypersetup{
    colorlinks=true,       
    linkcolor=black,       
    urlcolor=blue,         
    citecolor=black        
}
\usepackage{graphicx}
\usepackage{pdfpages}  
\ifCLASSINFOpdf
\else
\fi
\begin{document}
\title{Poster: Long PHP webshell files detection\\ based on sliding window attention}

\author{\IEEEauthorblockN{Zhiqiang Wang}
	\IEEEauthorblockA{Beijing Electronic Science\\ $\&$ Technology Institute, Beijing, China\\
		wangzq@besti.edu.cn}
	\and
	\IEEEauthorblockN{Haoyu Wang}
	\IEEEauthorblockA{Beijing Electronic Science\\ $\&$ Technology Institute, Beijing, China\\
		20232909@mail.besti.edu.cn}
	\and
	\IEEEauthorblockN{Lu Hao}
	\IEEEauthorblockA{Beijing Municipal Public\\ Security Bureau, Beijing, China\\
		hlucky@2008.sina.com}}

\maketitle

\begin{abstract}
Webshell is a type of backdoor, and web applications are widely exposed to webshell injection attacks. Therefore, it is important to study webshell detection techniques. In this study, we propose a webshell detection method. We first convert PHP source code to opcodes and then extract Opcode Double-Tuples (ODTs). Next, we combine CodeBert and FastText models for feature representation and classification. To address the challenge that deep learning methods have difficulty detecting long webshell files, we introduce a sliding window attention mechanism. This approach effectively captures malicious behavior within long files. Experimental results show that our method reaches high accuracy in webshell detection, solving the problem of traditional methods that struggle to address new webshell variants and anti-detection techniques.
\end{abstract}


% \fi
\IEEEpeerreviewmaketitle

\section{Introduction}
The webshell injection plays a vital role in the hacker attack chain, enabling the attacker to remotely control devices, acquire sensitive data, and further expand attack activities. Therefore, Detecting and removing webshells is an effective way to defend against attacks and ensure web security.

Traditional webshell detection methods \cite{p1,p2} based on pattern matching usually rely on recognizing known features, including source code features, traffic features, dynamic function calls and other relevant features. However, as attack techniques evolve, the variability and obfuscation of webshells have become more prevalent. Attackers often use obfuscation, dynamic loading, encryption and decryption techniques to evade detection, making traditional detection methods inadequate for recognizing new types of webshells.

In this context, webshell detection methods using deep learning \cite{p3,p4,p5}, including those based on source code or opcode, have become a research hotspot and have shown promising results. However, current deep learning-based webshell detection methods still face challenges \cite{p6}. For datasets, publicly available datasets are outdated and do not contain the latest samples. Therefore, their performance in real-world environments for detecting may not be good. For data processing, a good data processing method is often more important than the detection model. The opcode-based detection methods typically extract only a single sequence of opcode instructions (called Opcode Single-Tuples) without effectively capturing low-level code features. The source code-based method is complicated for processing webshells that use anti-detection techniques. In addition, detecting long sequence files (such as complex dynamic encryption and decryption scripts or large files) is quite challenging. Methods such as sample slicing \cite{p3} or TextRank \cite{p5} are often used to reduce data size, which may result in some loss of code information or disruption of contextual relationships.

This study focuses on the PHP language because PHP is used by 75.1$\%$ of all the websites whose server-side programming language \cite{p7}. To address the challenges, this study contribution includes (1) collating a new high-quality Webshell dataset, (2) proposing a PHP code data processing method to extract Opcode Double-Tuples(ODTs) including opcode instructions and operands instead of Opcode Single-Tuples(OSTs), (3) introducing a window attention mechanism to solve the long text problem.

% (should never be an issue

\section{METHODOLOGY}
The detection method consists of two steps. First, the PHP source code in the dataset is processed into ODTs. Second, using a sliding window attention mechanism, we combine the CodeBert model \cite{p8} and the Fasttext model \cite{p9} for feature representation and binary classification of the ODTs. Our dataset and processing code are publicly available: https://github.com/w-32768/PHP-Webshell-Detection-via-Opcode-Analysis

\begin{figure}[h]
\begin{center}
\includegraphics[width=2.8in,height=1.6in]{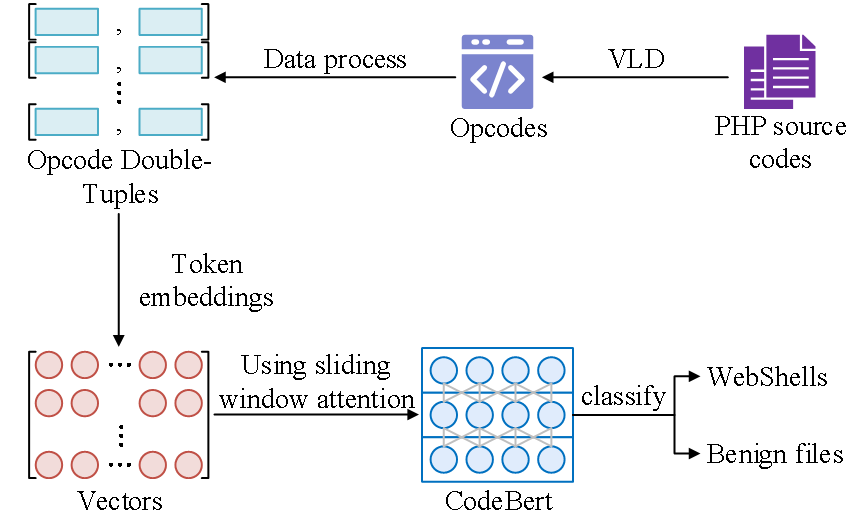}    % The printed column
\caption{Overview of the detection method.}  % width is 8.4 cm.
\label{fig1}                                 % Size the figures
\end{center}                                 % accordingly.
\end{figure}
\vspace{-0.8cm}
\subsection{Data processing}
The dataset consists of PHP source code files containing 5001 webshell samples and 5936 benign PHP files. Firstly, we convert the PHP source code to the opcode. The opcode, generated by the Zend Engine in PHP, is a low-level abstraction of source code. As anti-detection techniques are mostly used at the source code level, we have a natural advantage in using opcode detection.

After obtaining the opcodes, a series of data processing steps are performed. We use expert knowledge to establish fine-grained processing rules, extracting high-value instructions for detection while excluding those of low relevance, thus reducing opcode length without compromising contextual semantics. Operands may be encoded by URL or Base64 encoding, making it difficult to determine their semantics. Therefore, we perform the decoding operation. The original string content is restored based on string feature recognition. After this extraction, we have the set of opcode instructions and operands, called Opcode Double-Tuples. Experimental comparisons show that, under the same detection model training on our dataset, ODTs achieve a 4.6$\%$ accuracy improvement compared to OSTs, confirming that our data processing method is advanced and professional.

\subsection{Feature Representation and Binary Classification}
After data processing, this study explores using the CodeBert model and various embedding models for feature representation and binary classification of ODTs. The steps are as follows:

1) Feature Representation.
\begin{itemize}
\item \textbf{CodeBert Model:} The CodeBert Model is a widely used pre-trained language model optimized for code understanding tasks and pre-trained on PHP code. We input the ODTs into the CodeBert model to generate high-dimensional feature vector representations that capture the semantic and syntactic information of the opcodes.
\item \textbf{Embedding Models:} To enhance opcode feature representation, we compared four embedding models: Word2Vec, FastText, Glove, and Doc2Vec. Experimental comparisons show that FastText performs best in the opcode classification task; therefore, we chose FastText as the embedding model.
\item \textbf{Feature Fusion:} We fuse the feature vectors generated by CodeBert with the embedding vectors from FastText to form the final feature representation. The specific fusion formula is as follows:
    \begin{equation}
     E=\lambda E_{\rm CodeBert}+(1-\lambda)E_{\rm FastText}
     \vspace{-0.2cm}
    \end{equation}
$E_{\rm CodeBert}$ and $E_{\rm FastText}$ represent the feature vectors generated by CodeBert and FastText, respectively. $\lambda$ is the weight coefficient, and its optimal value is determined through experimentation.
\end{itemize}

2) Sliding Window Attention Mechanism:

We introduce a sliding window attention mechanism to address the high computational complexity of global self-attention mechanisms for long opcode sequences. The opcode sequence is divided into multiple windows of size $W$ with a stride of $Sr (Sr<W)$. Specifically, Self-attention is calculated independently within each window. The global feature representation is obtained by averaging the last hidden states from the CodeBert encoder across all windows. This mechanism reduces memory requirements and allows longer sequences to be processed. Furthermore, the overlap between adjacent windows allows information exchange, making it possible to detect malicious behaviors.

The sliding window attention mechanism reduces computational complexity and preserves the contextual information of the opcode sequence. Thus, the problem of incomplete information caused by other methods is avoided.

3) Binary Classification:

After getting the global feature representation of the ODTs, we input them into a binary classifier. The classifier consists of fully connected layers and activation functions, trained by minimizing the binary cross-entropy loss function. It distinguishes between benign PHP code and malicious webshells.

4) Model Training and Evaluation:

We fine-tuned the CodeBert model using the AdamW optimizer. Experimental results show that our proposed optimal model achieves an accuracy of 99.2$\%$ and an F1 score of 99.1$\%$ on the test set. Comparative experiments with accessible state-of-the-art webshell detection methods, including webshellPub \cite{p2} (Acc: 77.3$\%$, F1: 68.5$\%$), PHP Malware Finder \cite{p1} (Acc:83.4$\%$, F1:78.9$\%$), and MSDetector \cite{p3} (Acc:97.1$\%$, F1: 97.3$\%$), demonstrate the superiority of our method.

\section{CONCLUSION}
This study presents a PHP webshell data processing method that extracts ODTs, addressing the limitations of single-tuples detection. Additionally, we introduce a sliding window attention mechanism that effectively mitigates the challenges of long text detection. This study offers a new perspective on the field of malicious code detection. In the future, we aim to continually explore multi-language webshell detection tasks to improve detection performance and generalization capabilities.

\section*{Acknowledgment}
This work was supported by “the Fundamental Research Funds for the Central Universities” (Grant Number:3282024050).



%\bibliographystyle{IEEEtran}

\newpage
\includepdf[pages=-]{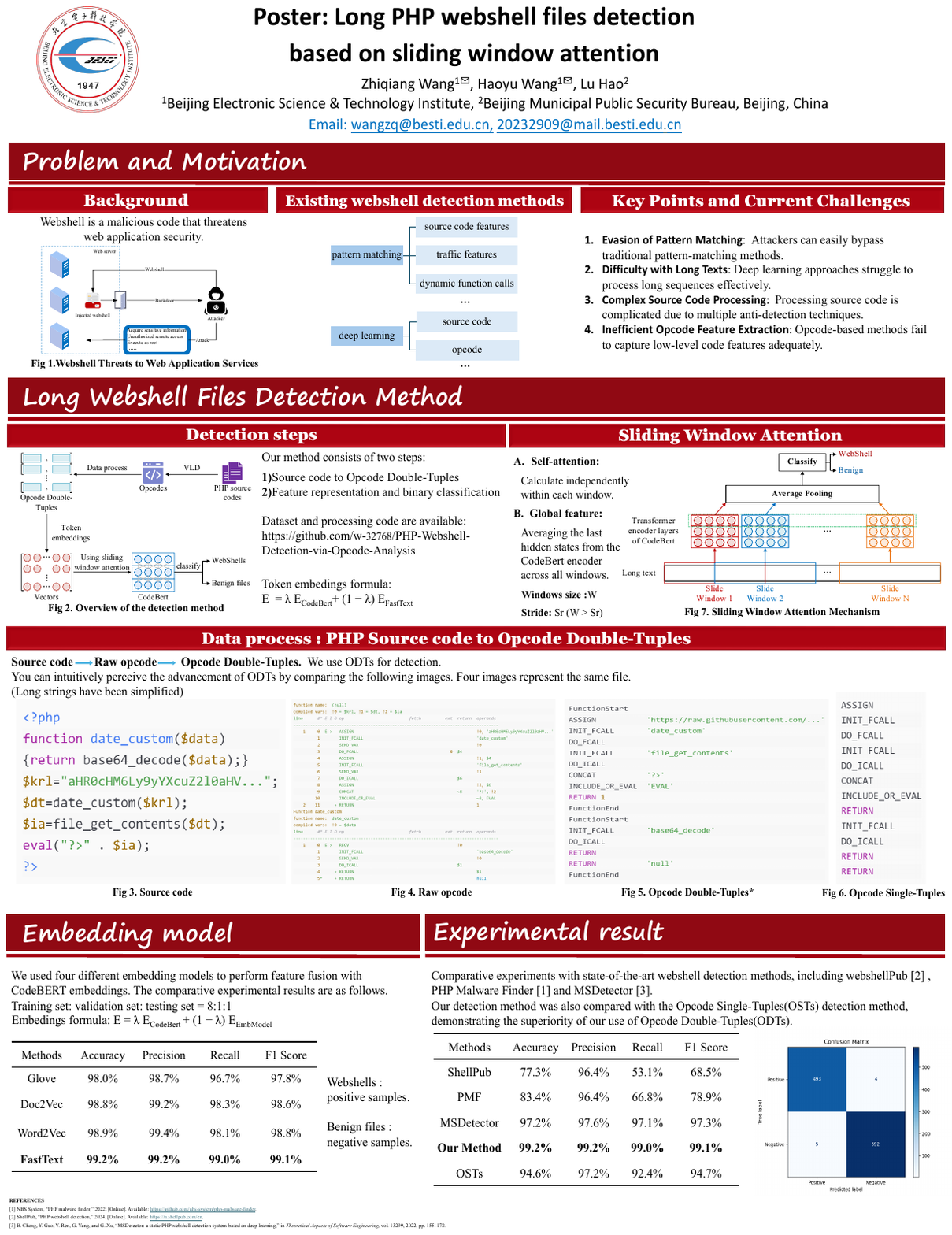}  % poster.pdf

\end{document}